\documentclass{article}

\usepackage{arxiv}

\usepackage[utf8]{inputenc} 
\usepackage[T1]{fontenc}    
\usepackage{hyperref}       
\usepackage{url}            
\usepackage{booktabs}       
\usepackage{amsfonts,bm}       
\usepackage{nicefrac}       
\usepackage{microtype}      
\usepackage{lipsum}		
\usepackage{graphicx}
\usepackage{natbib}
\usepackage{doi}

\title{Accelerated 3D Electrical Resistivity Tomography with a Scalable Jacobian-free Approach}

\author{
  \href{https://orcid.org/0000-0002-3646-2915 }{\includegraphics[scale=0.06]{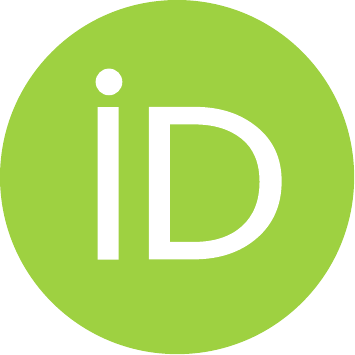}\hspace{1mm}Jonghyun Lee}\\
  University of Hawai'i at Man\=oa \\
  Honolulu, HI 96822 \\
  \texttt{jonghyun.harry.lee@hawaii.edu} \\
 }

\renewcommand{\shorttitle}{3D ERT with a Scalable Jacobian-free Approach}

\hypersetup{
pdftitle={Accelerated 3D Electrical Resistivity Tomography with a Scalable Jacobian-free Approach},
pdfsubject={physics.geo-ph},
pdfauthor={Jonghyun Lee},
pdfkeywords={Principal Component Geostatistical Approach, Electrical Resistivity Tomography, 3D Subsurface Characterization},
}

\begin{document}
\maketitle

\begin{abstract}
A Jacobian-free inversion method is presented to accelerate Electrical Resistivity Tomography (ERT) for shallow aquifer characterization. The ERT problem typically implements the adjoint state method to efficiently compute Jacobian during the inversion. However, the adjoint state method needs intrusive forward model code changes and may not be computationally scalable with many observations especially when one performs 3D ERT surveys with dense multi-electrode arrays. Here the Principal Component Geostatistical Approach (PCGA), a fast and scalable Jacobian-free inverse modeling method, is applied to solve a high dimensional data-intensive ERT problem. The PNNL’s ERT simulation software E4D was linked to the python interface pyPCGA without intrusive code change and the example code is upload in a public repository. The result in this study shows that high-resolution 3D subsurface characterization is computationally feasible, which would have a great potential for implementations in practice.
\end{abstract}


\section{Introduction}
Characterization of geologic heterogeneity is crucial for subsurface management operations such as aquifer recharge and recovery or remediation of organic contaminants. With recent advances in computational resources and sensor technology, high-resolution subsurface images can be achieved using various geophysical surveys. However, such inversion with a large volume of geophysical measurements requires high, often prohibitive, computational costs associated with a number of large-scale numerical simulation runs and high-dimensional dense matrix multiplications. As a result, traditional inversion techniques have limited utility for the high-dimension and/or data-intensive problems, e.g., applications that require fine discretization on large domains and many measurements to capture small-scale subsurface heterogeneity, like gas pipeline leakage detection or preferential flow of contaminants~\citep{ghorbanidehno2020recent}.

In this work, we present an efficient inversion method for 3D Electrical Resistivity Tomography (ERT) applications.  The domain we consider is a synthetic three-dimensional shallow aquifer with 250,000 unknown conductivities. 65,000 potential measurements from dipole-dipole tests are used to delineate the heterogeneous conductivity field and to quantify the corresponding estimation uncertainty. For this high-dimensional data-intensive application, the Principal Component Geostatistical Approach (PCGA)~\citep{lee2014large}, a computationally efficient and scalable Jacobian-free inverse modeling approach, is applied. PCGA can dramatically reduce the computational costs of the inversion by using a low rank approximation of the high dimensional prior covariance matrix followed by the sensitivity computation along the dominant principal component directions, effectively reducing the number of required forward simulations by orders of magnitude. PCGA utilizes forward simulators as a black box, which facilitates a seamless and flexible integration with any geophysics modeling software without intrusive code changes. In addition, the method can harness the parallel capabilities of available computer resources to efficiently scale to large-scale problems. 

\section{Method}
\label{sec:method}

The objective of the inverse modeling or variational data assimilation is to determine a vector of unknowns from a vector of observations. Due to the uncertainty in data collection as well as physics modeling, the Bayesian geostatistical inverse approach~\citep{kitanidis1995quasi}, an optimization-based stochastic inverse modeling in a Hierarchical Bayesian framework~\citep{kitanidis2011bayesian} is used in this study. The observation equation is given by:

\begin{equation}
    \mathbf{d} = \mathbf{g}(\mathbf{m}) + \bm{\varepsilon},\;\; \bm{\varepsilon} \sim \mathcal{N}(0, \mathbf{C}_{\mathbf{d}}^{prior})
\end{equation}
where $\mathbf{d}$ is the observation data such as observed potential measurements, $\mathbf{g}$ is a forward model such as any ERT simulator, $\mathbf{m}$ is the unknown spatially varying electric conductivity field, and $\bm{\varepsilon}$ is the error in the observation data $\mathbf{d}$ as well as the simulation model $\mathbf{g}$, usually modeled as Gaussian after power transformation. To estimate unknown $\mathbf{m}$ from the data and the forward model, In the Hierarchical Bayesian framework, the prior probability of $\mathbf{m}$ is assumed to be Gaussian with an unknown mean and a prior covariance matrix $\mathbf{C}_\mathbf{m}^{prior}$. Then, the posterior pdf of m is computed through Bayes' theorem and the maximum a posteriori (MAP) estimate or most likely value of m is obtained by maximizing the posterior pdf (typically minimizing the negative log likelihood of the posterior pdf). The posterior uncertainty is quantified through linearization around the MAP estimate. The inverse problem becomes a nonlinear optimization problem that is commonly solved using iterative Gauss-Newton method. While the geostatistical approach is well suited for small- to moderate scale inverse problems, it becomes computationally challenging for large-scale data-intensive inversions because of the computation of the derivative of the forward model, i.e., Jacobian or sensitivity matrix $\mathbf{J}$ at a current estimate $\tilde{\mathbf{m}}$:
\begin{equation}
    \mathbf{J} = \left.\frac{\partial \mathbf{g}}{\partial \mathbf{m}} \right|_{m = \tilde{m}}
\end{equation}
The Jacobian matrix is not typically implemented alone but multiplied by the prior covariance matrix to yield cross-covariance $\mathbf{C}_{\mathbf{dm}}$  between the data d and the unknowns m and data covariance $\mathbf{C}_{\mathbf{dd}}$:

\begin{equation}
    \label{eq:cross_cov}
    \mathbf{C}_{\mathbf{dm}} = \mathbf{J}\mathbf{C}_\mathbf{m}^{prior},\;\; \mathbf{C}_{\mathbf{dd}} = \mathbf{J}\mathbf{C}_\mathbf{m}^{prior}\mathbf{J}^{\top}
\end{equation}
The construction of these Jacobian-covariance products is the main computational bottleneck in the inversion. The direct construction of Jacobian requires computational storage and costs proportional to the number of observations, even in the efficient adjoint-state method. The dense matrix multiplication in Equation~\ref{eq:cross_cov} is challenging even in the high performance computing environment. Furthermore, the adjoint-state method needs intrusive changes in the forward model code, which make the model developer challenging. 

To address the computational issues, indirect approaches such as Conjugate Gradient-Gauss Newton (CG-GN) method have been introduced to avoid the direct construction of Jacobian, however, the CG-GN methods require many \emph{sequential} CG inner iterations on top of outer sequential GN iterations which cannot be parallelized on multicore computers. In addition, the second order adjoint-state method for the inner CG iterations needs to be implemented using the current forward modeling code. 

\subsection{Principal Component Geostatistical Approach (PCGA)}
PCGA expedites the stochastic inversion by avoiding the direct evaluation of the Jacobian matrix. It utilizes a low-rank approximation of the covariance $\mathbf{C}_m$ and a finite difference approximation of all matrix-vector products. Through the truncated eigendecomposition, $\mathbf{C}_m$ is approximated through:
\begin{equation}
    \label{eq:cov_approx}
    \mathbf{C}_\mathbf{m} \approx \mathbf{C}_{n_{pc}} = \sum_{i=1}^{n_{pc}} \bm{\zeta}_i \bm{\zeta}_i^{\top}
\end{equation}
where $\mathbf{C}_{n_{pc}}$ is a rank-$n_{pc}$ approximation of $\mathbf{C}_{m}$, $\zeta_i$ is i-th eigenvector multiplied by square root of i-th eigenvalue of $\mathbf{C}_m$. A fast method to obtain Eq. (3) for large-scale covariance matrices is explained in~\citep{wang2021pbbfmm3d}. 
\begin{equation}
    \label{eq:crosscov_approx}
    \mathbf{C}_{\mathbf{dm}} = \mathbf{J} \mathbf{C}_\mathbf{m} \approx \mathbf{J} \left( \bm{\zeta}_i \bm{\zeta}_i^{\top}\right) = \sum_{i=1}^{n_{pc}} \left( \mathbf{J} \bm{\zeta}_i \right) \bm{\zeta}_i^{\top}
\end{equation}
where $\mathbf{J}\bm{\zeta}_i$ is computed using finite-difference with a small perturbation $\delta$ (e.g., square root of machine precision) as 
\begin{equation}
    \label{eq:fd_approx}
    \mathbf{J}{\bm{\zeta}_i} \approx \frac{\mathbf{g}(\mathbf{m}+\delta \bm{\zeta}_i) - \mathbf{g}(\mathbf{m})}{\delta}
\end{equation}
Through a series of approximation in Equations~\ref{eq:cov_approx}-~\ref{eq:fd_approx}, $n_{pc}+1$ forward simulation evaluations at each iteration are needed to obtain the inverse solution. The number of numerical simulations can be reduced by a factor of 10 or more with controlled accuracy in most cases~\citep{lee2014large,lee2016scalable,kang2020improved}. In practice, PCGA can reduce the number of numerical forward simulations greatly, usually up to a few hundred simulation runs, which is highly beneficial for complex transient joint inversion problems, while inverse solutions are almost the same as those obtained from the conventional geostatistical approach. To illustrate the efficiency and scalability of PCGA, an ERT problem is presented in this work.

\section{Application}
\label{sec:application}
\begin{figure}
	\centering
	\includegraphics[width=0.8\textwidth]{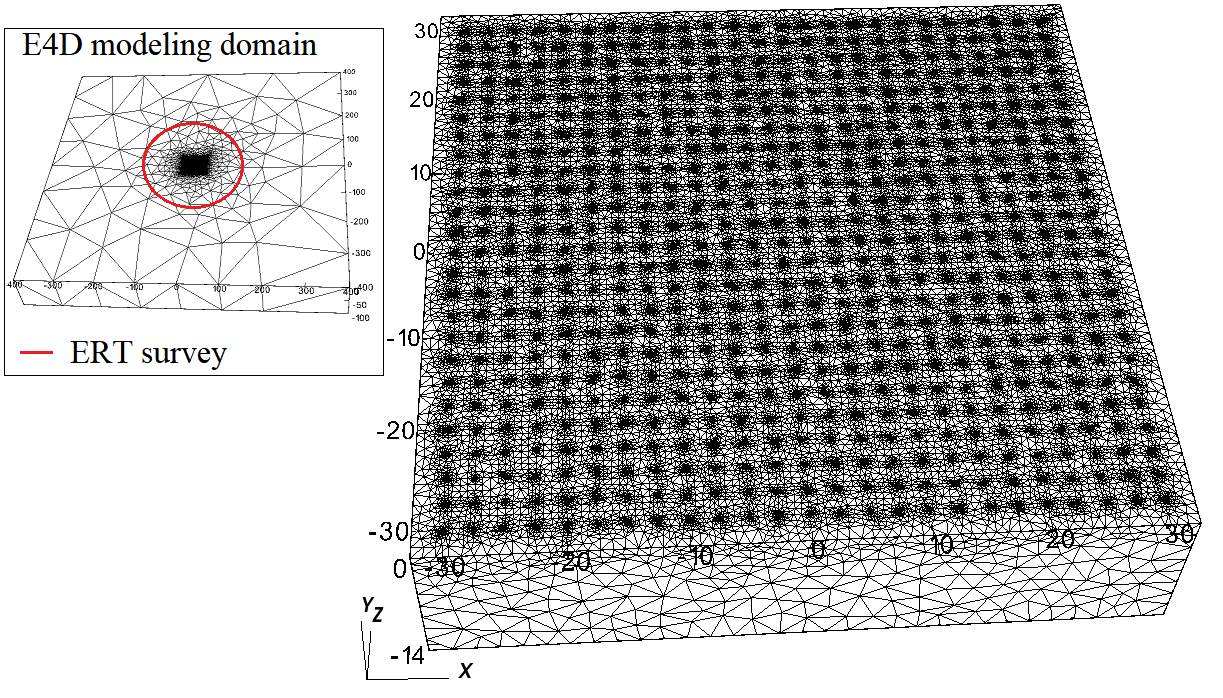}
	\caption{E4D finite element mesh configuration over the entire 800 m x 800 m x 100 m domain (inset) and 60 m x 60 m x 30 m ERT survey volume (left).}
	\label{fig:fig1}
\end{figure}

To illustrate the scalability of the proposed method, a 60 m x 60 m x 30 m (width x length x depth) shallow aquifer is characterized with a ERT survey as shown in Figure~\ref{fig:fig1}. ERT forward simulation is performed with PNNL’s E4D simulator~\citep{johnson2010improved}. The 3D survey array consists of 25 parallel 25-electrode arrays (i.e., 625 electrodes in total) with 2.4 m spacing installed on the surface of the aquifer. A series of inline and cross-line dipole-dipole potential measurements was collected. To eliminate the effect of boundary conditions in the forward modeling and inversion, the size of ERT simulation domain is set to 800 m x 800 m x 100 m as shown in the inset of Figure~\ref{fig:fig1}. The discretization of the domain is fine close to the center with mesh refinement performed at the 625 electrodes and it becomes coarser with distance from the domain of interest. The fine discretization was applied to test the feasibility of small-scale subsurface feature detection and illustrate the computation scalability of the method. The total number of finite elements resulting from this discretization is about 500,000.

\begin{figure}
	\centering
	\includegraphics[width=1.0\textwidth]{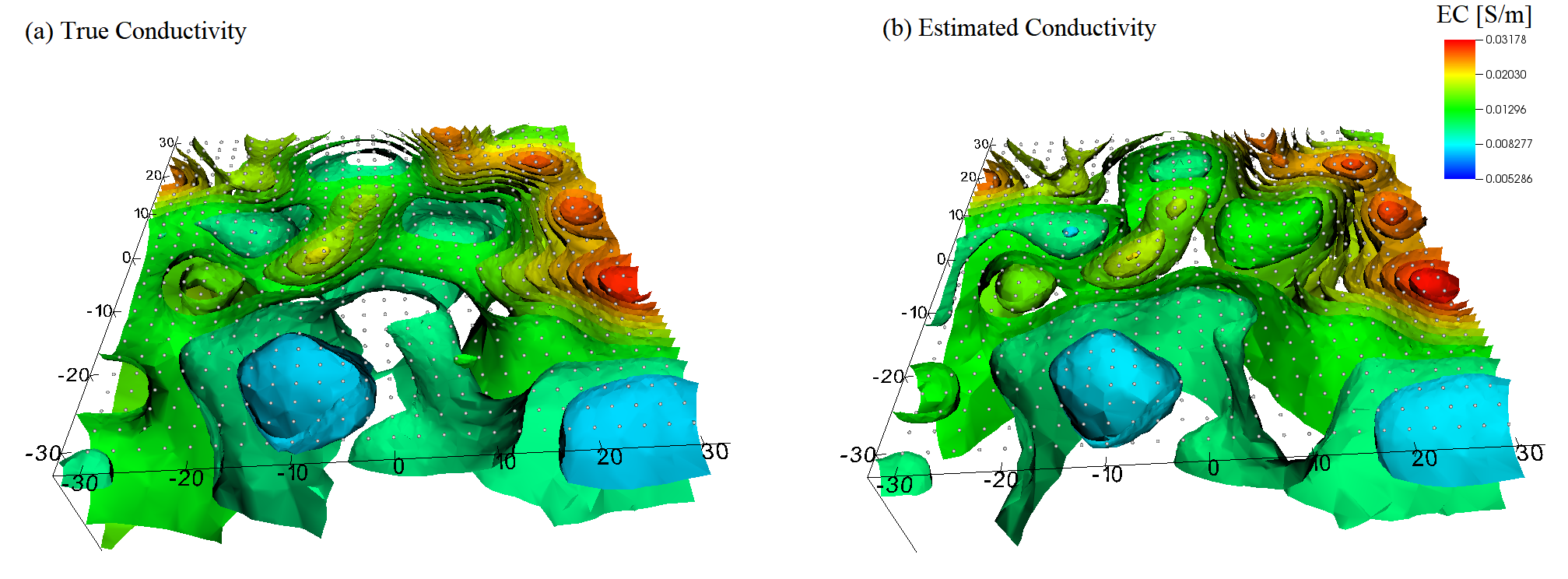}
	\caption{(a) Synthetic true electrical conductivity distribution [S/m]. (b) Estimated electrical conductivity distribution [S/m].}
	\label{fig:fig2}
\end{figure}

The true electrical conductivity field in~\ref{fig:fig2}(a) was generated from lognormal distribution with the mean of 0.1 S/m and an exponential covariance with scale parameter $lx$ = 30 m, $ly$ = 30 m, $lz$ = 20 m. With the true conductivity field, potential measurements were simulated and both 5\% transfer resistance measurement error and random noise were added to the measurements. For illustration purposes, only 25 potential measurements from each current electrode pair were considered resulting in 60,235 measurements in total. Inclusion of more observations did not change the estimate due to weak sensitivity of the dipole-dipole survey measurements at distant locations.

Since we expect that the estimated electrical conductivities outside the E4D survey network would not be improved from the prior, only the center volume of 90 m x 90 m x 30 m are characterized in this application and the outside electrical conductivity is assigned to the mean of the current estimate. The number of unknown electrical conductivities is set to 256,111 defined on the uniform grids with 1 m spacing (nx = 91, ny = 91, nz = 31). In the inversion, the conductivity field is interpolated on the E4D’s finite element grids to simulate potential measurements. Still, reconstructing a half million electric conductivity values on unstructured grids directly is feasible with the proposed approach~\citep[e.g.][]{wang2021pbbfmm3d} and will be presented elsewhere. Tables~\ref{tab:problem} and~\ref{tab:inversion} present forward model and inversion parameters, respectively.

\begin{table}
	\caption{Problem Setting}
	\centering
	\begin{tabular}{ll}
		\toprule
		Parameter    & Value\\
		\midrule
		Domain (x,y,z) & 800, 800, 300 (m)  \\
		\# of finite elements     & 499,964 \\
		\# of unknowns     & 256,711  \\
		\# of electrodes     & 625 (25 lines x 25 electrodes)  \\
		\# of observations    & 65,235  \\
		\# of unknowns     & 15  \\
		\bottomrule
	\end{tabular}
	\label{tab:problem}
\end{table}

\begin{table}
	\caption{Inversion parameters}
	\centering
	\begin{tabular}{ll}
		\toprule
		Parameter    & Value\\
		\midrule
		Covariance function & Exponential \\
		Prior variance & $0.5^2$ (log(S/m)$^2$)\\
		Correlation length (lx,ly,lz) & 30, 30, 15 (m)  \\
		$n_{pc}$ & 96 \\
		Total \# of E4D runs & 420\\
		Inversion time     & 6 (hours)  \\
		\bottomrule
	\end{tabular}
	\label{tab:inversion}
\end{table}
Figure~\ref{fig:fig2}(b) shows the best estimate. Since a large number of survey data is used, small-scale features of the true field (Figure~\ref{fig:fig2}(a)) are identified accurately especially in the shallow part of the aquifer. The estimated conductivity in the deeper region becomes less accurate due to weak sensitivity in the measurements as expected. The estimation uncertainty in Figure 4 also confirms the posterior standard deviation at a depth of 15 m below the surface is close to the prior variance meaning that the information from the survey was not enough to reduce the uncertainty significantly. 

The best estimate converged in 4 iterations with 420 simulations in total in 6 hours on an Intel 48-core 2.1 GHz workstation and only 2 GB RAM was required. The promising computational efficiency and scalability in the proposed method implies that large-scale subsurface characterization with uncertainty quantification can be completed on the same day so that field practitioners can design subsequent surveys on a modern personal computer in the next days. Conventional inversion techniques would require much larger computational efforts and memory footprints for the ERT application, for example, a similar problem was solved on distributed 400 processors in 5 hours with $\sim$300 GB RAM~\citep{johnson2010improved}. 

Note that the E4D simulator was used as a black box and only a python wrapper to link E4D with pyPCGA by updating electrical conductivities and reading simulation outputs was required for the inverse modeling. Example code used in this study can be found in the link below:
\url{https://github.com/jonghyunharrylee/pyPCGA/tree/master/examples/ERT_E4D}
 
\section{Conclusion}
A Jacobian-free stochastic inversion method called PCGA is used to perform 3D ERT with a number of potential measurements and provided reasonable inversion results with affordable E4D runs. PCGA transforms an inverse problem with the computational cost associated with the number of observations into an approximately same problem with a constant number (i.e., total O(100)) of simulations, so that one would expect a great computational gain in solving large-scale inverse problems without the adjoint-state method implementation. The application in this paper implies that large-scale inversion at real field sites can become computationally feasible and have a great potential in the field studies. 

\section{Acknowledgements}
E4D and its can be downloaded from \url{https://e4d.pnnl.gov}. The python package pyPCGA can be downloaded from \url{https://github.com/jonghyunharrylee/pyPCGA}. The ERT inversion example can be downloaded from  \url{https://github.com/jonghyunharrylee/pyPCGA/tree/master/examples/ERT_E4D}.









\bibliographystyle{unsrtnat}
\bibliography{references}  






\end{document}